\documentclass{ws-mpla}
\newcommand{\PRL}{Phys. Rev. Lett. }
\newcommand{\PR}{Phys. Rev. }
\newcommand{\PL}{Phys. Lett. }
\newcommand{\CQG}{Class. Quantum Grav. }
\newcommand{\APJ}{Astrophys. J. }
\newcommand{\NP}{Nucl. Phys. }
\newcommand{\be}{\begin{equation}}
\newcommand{\ee}{\end{equation}}
\begin{document}
\title{A new alternative model to dark energy}
\author{Yungui Gong}
\address{Institute of Applied Physics and College of Electronic
Engineering\\
Chongqing University of Post and Communication\\
Chongqing 400065, China\\ gongyg@cqupt.edu.cn}
\author{Xi-Ming Chen}
\address{College of Electronic
Engineering\\ Chongqing University of Post and Communication\\
Chongqing 400065, China\\
chenxm@cqupt.edu.cn}
\author{Chang-Kui Duan}
\address{Institute of Applied Physics and College of Electronic
Engineering\\ Chongqing University of Post and Communication\\
Chongqing 400065, China\\ duanck@cqupt.edu.cn}
\maketitle
\begin{abstract}
The recent observations of type Ia supernovae strongly support
that the universe is accelerating now and decelerated in the
recent past. This may be the evidence of the breakdown of the
standard Friedmann equation. Instead of a linear function of the
matter density, we consider a general function of the matter
density to modify the Freidmann equation. We propose a new model
which explains the recent acceleration and the past deceleration.
Furthermore, the new model also gives a decelerated universe in
the future. The new model gives $\Omega_{\rm m0}=0.46$ and $z_{\rm
T}=0.44$.
\keywords{dark energy; alternative model; Type Ia
supernova.}
\end{abstract}
\ccode{PACS Nos.: 98.80.-k, 98.80.Es,04.50.+h,14.80.-j}

The possible discovery of an accelerating universe from
observations of Type Ia supernovae (SNe Ia) leads to a new wave of
interest in cosmology \cite{sp99}-\cite{agriess}. Observational
results also provide the evidence of a decelerated universe in the
recent past \cite{agr,mstagr}. On the other hand, the cosmic
microwave background (CMB) observations indicate that the universe
is spatially flat as predicted by inflationary models
\cite{pdb00,bennett03}. These observations suggest that the
universe is dominated by dark energy. One simple candidate of dark
energy is the cosmological constant. However, the unusual small
value of the cosmological constant leads to the search for
dynamical dark energy models although cold dark matter
cosmological constant ($\Lambda$-CDM) models are consistent with
the current observations. The quintessence model is one of the
alternatives \cite{dark}-\cite{tachyon}. But there is no direct
evidence of dark energy and the nature of dark energy remains
mysterious. One logical alternative is that the standard Friedmann
equation may need to be modified. In this alternative scenario,
the universe is just dominated by ordinary pressureless matter,
but the laws of gravity and the standard Friedmann equations are
modified. The idea of modifying the laws of gravity is not new.
The modified Newtonian Dynamics (MOND) was first used to explain
the rotation curve in place of the dark matter \cite{mond,mond1}.
In MOND, the Newtonian gravity $M/r^2$ is replaced with
$M/r^2+\sqrt{M}/r$. Since $M/r^2$ gives the standard Freidmann
equation $H^2\sim \rho$, $M/r^2+\sqrt{M}/r$ may provide a modified
Friedmann equation $H^2\sim \rho^{2/3}\ln \rho+ \rho^{2/3}$
\cite{mond1}. Recall that the brane cosmology gives a non-standard
Friedmann equation $H^2\sim \rho +\rho^2$ \cite{brane}. Along this
line of reasoning, Freese and Lewis recently proposed the
Cardassian expansion in which the universe is dominated by the
ordinary matter and the Friedmann equation becomes $H^2\sim
\rho+\rho^n$ \cite{chung,freese02}. The Cardassian expansion model
was later generalized to a more general Friedmann equation
$H^2\sim g(\rho)$ \cite{chung}-\cite{gong03}. In addition, several
authors modified the Friedmann equation as $H^2+H^\alpha \sim
\rho$ motivated by theories with extra dimensions \cite{dvali}.
Furthermore, Chung and Freese argued that almost any relationship
between $H$ and $\rho$ is possible if our universe is a three
brane embedded in five dimensional spacetime \cite{chung}. In this
paper, we discuss an extra-dimension inspired model with
generalized Friedmann equation. We first review three models
analyzed by Gong and Duan \cite{gong03}: a model which is
equivalent to the generalized Chaplygin gas model \cite{chaply} in
terms of dynamical evolution, the generalized Cardassian model,
and the DGP model \cite{dvali}. Then we propose our new model and
explore the property of this new model.

For a spatially flat, isotropic and homogeneous universe with both
an ordinary pressureless dust matter and a minimally coupled
scalar field $Q$ sources, the Friedmann equations are \be
\label{cos1} H^2=\left({\dot{a}\over a}\right)^2={8\pi G\over
3}(\rho_{\rm m}+\rho_{\rm Q}), \ee \be \label{cos2} {\ddot{a}\over
a}=-{4\pi G\over 3}(\rho_{\rm m}+\rho_{\rm Q}+3p_{\rm Q}), \ee \be
\label{cos3} \dot{\rho_{\rm Q}}+3H(\rho_{\rm Q}+p_{\rm Q})=0, \ee
where dot means derivative with respect to time, $\rho_{\rm
m}=\rho_{\rm m0}(a_0/a)^3$ is the matter energy density, a
subscript 0 means the value of the variable at present time,
$\rho_{\rm Q}=\dot{Q}^2/2+V(Q)$, $p_{\rm Q}=\dot{Q}^2/2-V(Q)$ and
$V(Q)$ is the potential of the quintessence field. The modified
Friedmann equations for a spatially flat universe are \be
\label{cosa} H^2=H_0^2g(x), \ee \be \label{cosb} {\ddot{a}\over
a}=H^2_0g(x) -{3H^2_0x\over 2}g'(x)\left({\rho+p\over
\rho}\right),\ee \be \label{cosc} \dot{\rho}+3H(\rho+p)=0, \ee
where $x=\Omega_{\rm m}=8\pi G\rho/3H^2_0=x_0(1+z)^3$ during the
matter dominated epoch, $1+z=a_0/a$ is the redshift parameter,
$g(x)=x+\cdots$ is a general function of $x$ and $g'(x)=dg(x)/dx$.
Note that the universe did not start to accelerate when the other
nonlinear terms in $g(x)$ started to dominate. For the matter
dominated flat universe, $\rho=\rho_{\rm m}$ and $p=p_{\rm m}=0$,
we have $x_0=\Omega_{\rm m0}$, $g(x_0)=1$ and $x=\Omega_{\rm
m0}(1+z)^3$. To compare the modified model with dark energy model,
we make the following identification by using Eqs.
(\ref{cos1})-(\ref{cosc})
\begin{equation}
\label{darkom} \omega_{\rm Q}={xg'(x)-g(x)\over g(x)-x}.
\end{equation}
In general, $\omega_{\rm Q}$ is not a constant. The transition
from deceleration to acceleration happens when the deceleration
parameter $q=-\ddot{a}/aH^2=0$. From Eqs. (\ref{cosa}) and
(\ref{cosb}), we have \be \label{trans} g[\Omega_{\rm
m0}(1+z_{q=0})^3]={3\over 2}\Omega_{\rm
m0}(1+z_{q=0})^3g'[\Omega_{\rm m0}(1+z_{q=0})^3],\ee\be
\label{qparam} q_0={3\over 2}\Omega_{\rm m0}g'(\Omega_{\rm m0})-1.
\ee

Gong and Duan analyzed three models by using the WMAP and
supernovae data \cite{gong03}. The first model is
$$g(x)=x+(1-\Omega_{\rm m0})[A_s+(1-A_s)(x/\Omega_{\rm m0})^\beta]^{1/\beta},$$
which is equivalent to the generalized Chaplygin gas model
$p_c=-A/\rho_c^\alpha$ as dark energy with $\beta=1+\alpha$ and
$A_s=(8\pi G/3H^2_0(1-\Omega_{\rm m0}))^\beta A$ in terms of the
dynamical evolution of the universe.  The second model is the
generalized Cardassian model
$$g(x)=x[1+Bx^{\alpha(n-1)}]^{1/\alpha},$$
where $B=(\Omega_{\rm m0}^{-\alpha}-1)/\Omega_{\rm
m0}^{\alpha(n-1)}$, $\alpha>0$ and $n<1-1/3(1-\Omega^\alpha_{\rm
m0})$. The third model is the DGP model $g(x)=[a+\sqrt{a^2+x}]^2$,
where $a=(1-\Omega_{\rm m0})/2$.

Now let us consider a new extra-dimension inspired model
$g(x)=x(1+e^{-\alpha x})^n$ with $\alpha=-\ln(\Omega_{\rm
m0}^{-1/n}-1)/\Omega_{\rm m0}>0$. At high redshift $z$, $x$ is
very large and $g(x)\sim x$. Therefore the standard model is
recovered at early times. During the matter dominated epoch,
$p=0$. Eq. (\ref{cosb}) gives
\begin{equation}
\label{qzrel} {\ddot{a}\over a H^2_0}=-{1\over 2}x(1+e^{-\alpha
x})^n+{3\over 2}n\alpha x^2(1+e^{-\alpha x})^{n-1}e^{-\alpha x}.
\end{equation}
To have a clear picture of this model, we plot the above function
for different values of $\alpha$ and $n$.
\begin{figure}[htb]
\begin{center}
$\begin{array}{cc} \psfig{file=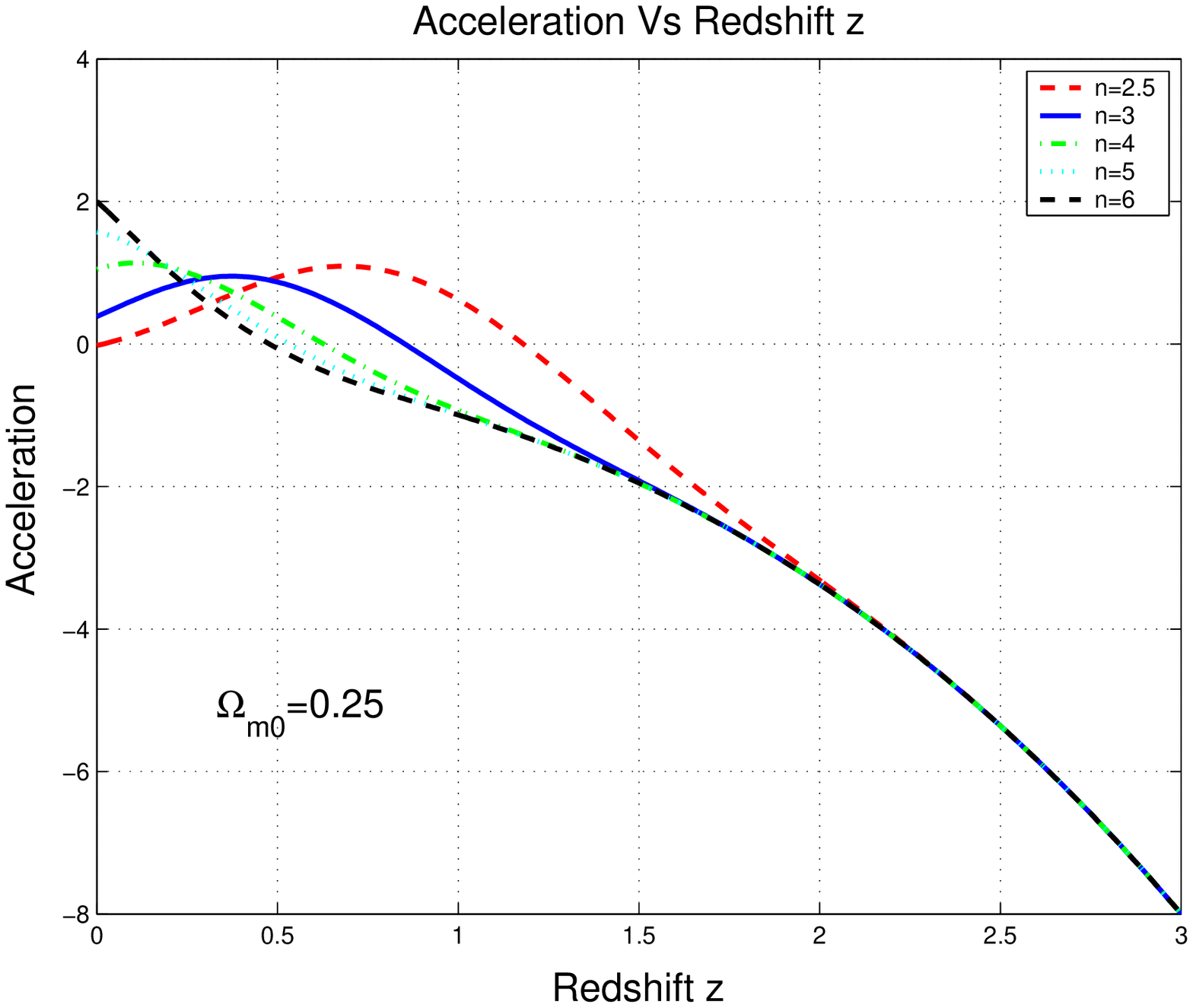,width=2.5in} &
\psfig{file=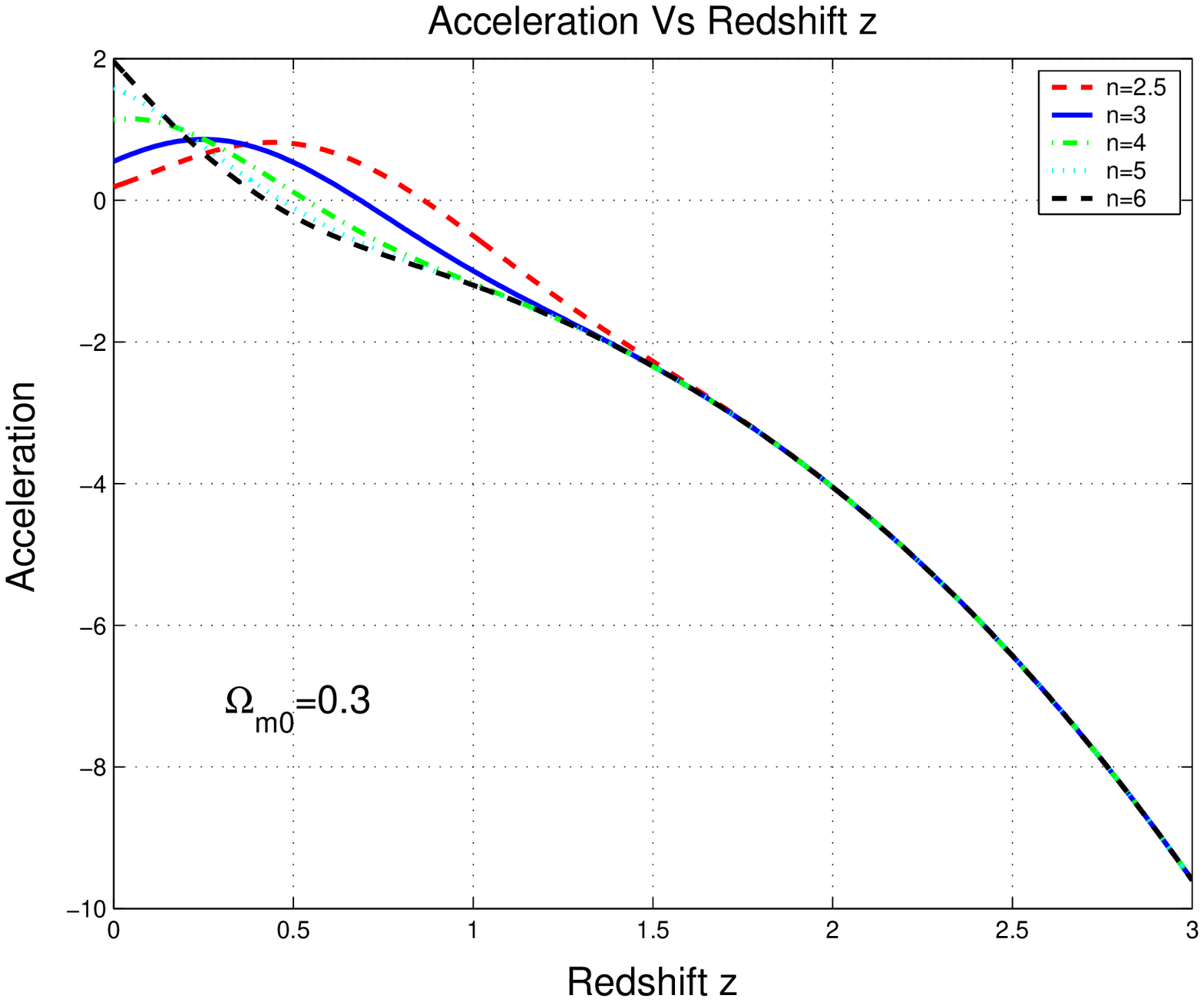,width=2.5in}
\end{array}$
\end{center}
\vspace*{8pt} \caption{ \label{qzplot} $\ddot{a}/aH^2_0$ versus
the redshift $z$ for different $\Omega_{\rm m0}$ and $n$ }
\end{figure}
From figure \ref{qzplot}, it is clear that the expansion of the
universe slowed down until the recent past, then the expansion
speeds up. However, there is an important new feature of this
model, the universe will not expand with eternal acceleration.
Therefore there is no future event horizon problem. We can also
see from figure \ref{qzplot} that the standard expansion recovered
at around $z\sim 2$. If we would like to make a comparison with
dark energy model, then we get from Eq. (\ref{darkom})
$$\omega_{\rm Q0}={n(1-\Omega_{\rm m0}^{1/n})\ln(\Omega_{\rm m0}^{-1/n}-1)\over
\Omega_{\rm m0}(1-\Omega_{\rm m0})}.$$

\begin{figure}[htb]
\begin{center}
$\begin{array}{cc} \psfig{file=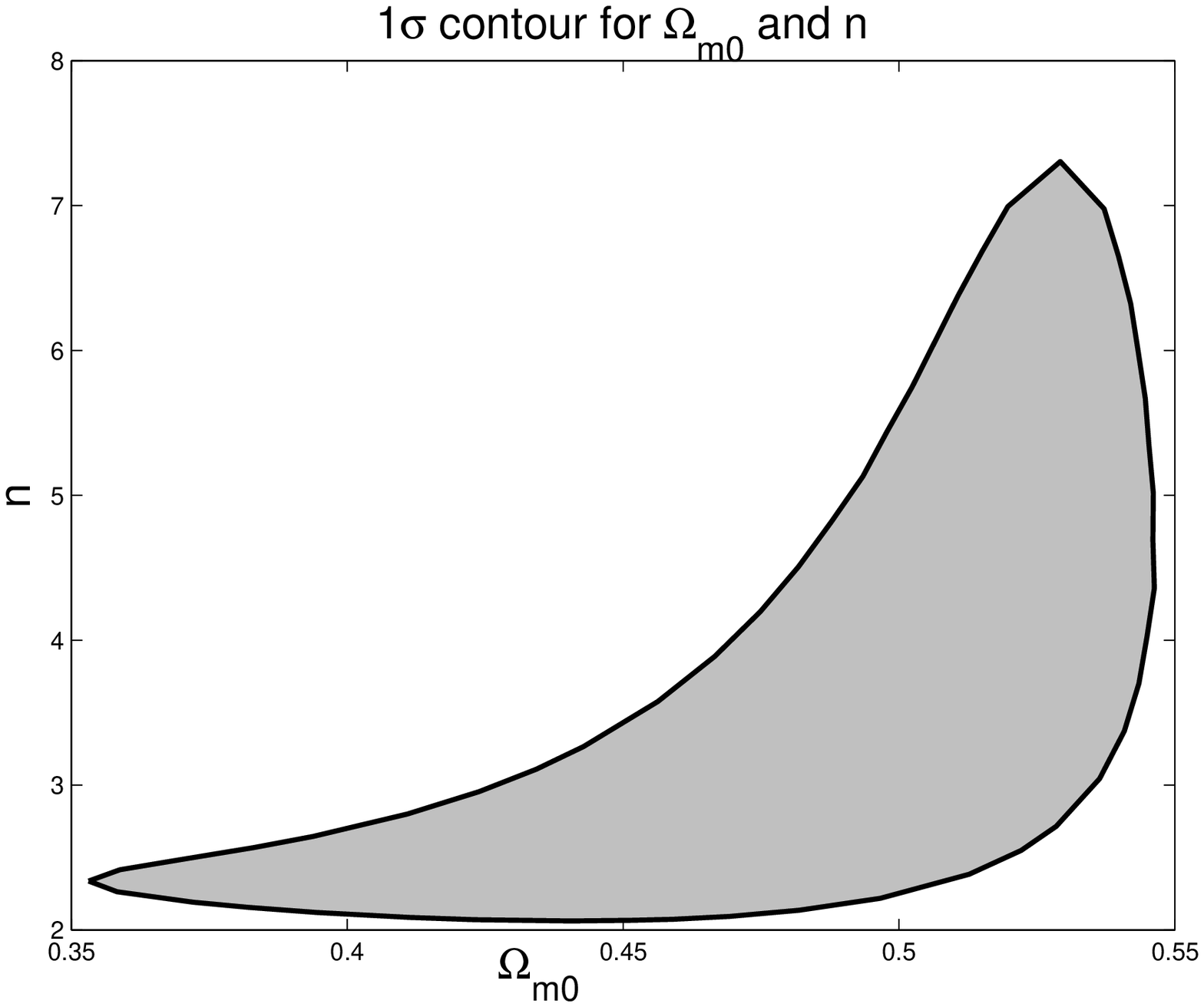,width=2.5in} &
\psfig{file=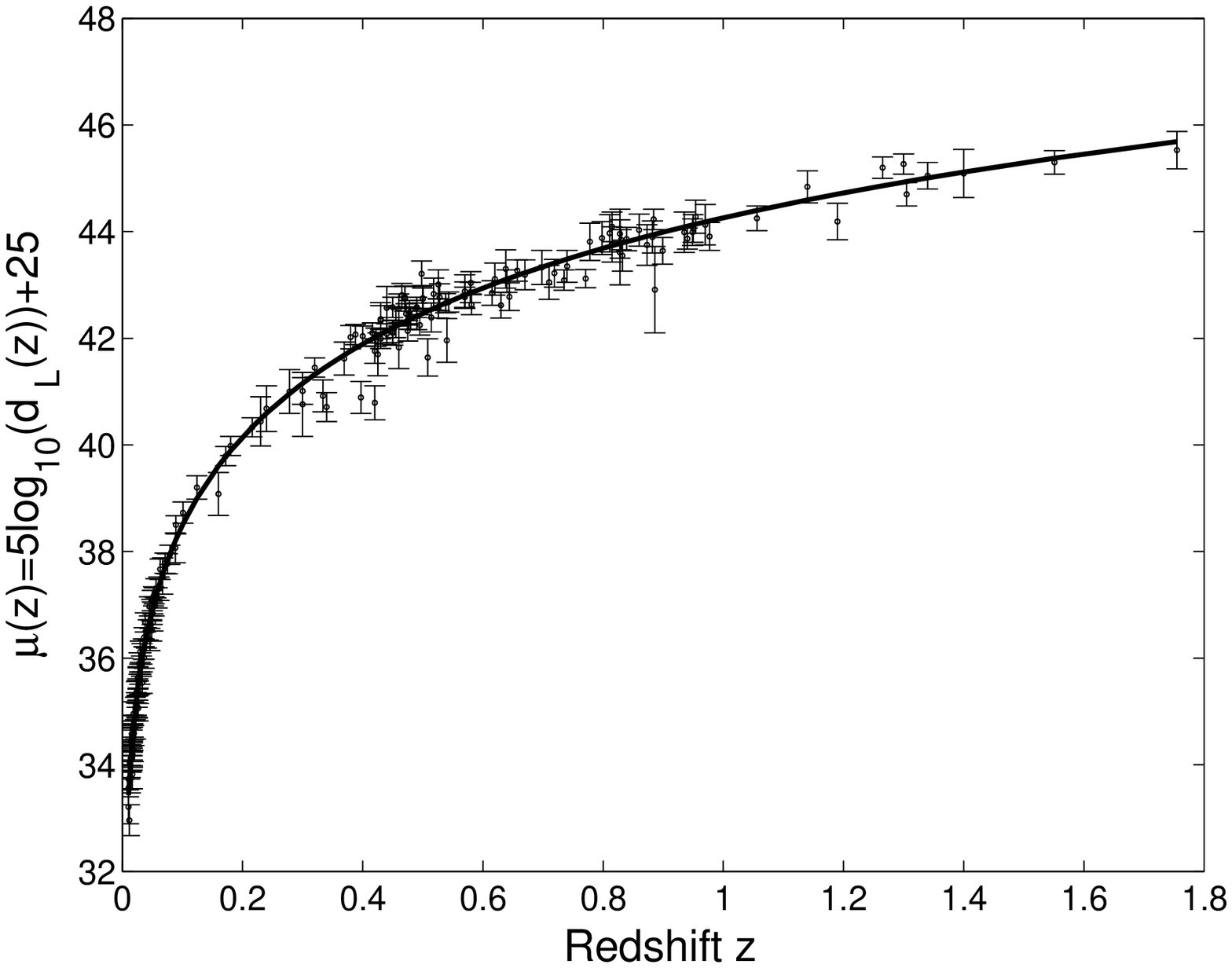,width=2.5in}
\end{array}$
\end{center}
\vspace*{8pt} \caption{ \label{bestfit} The left panel shows the
$1\sigma$ confidence contour for $\Omega_{\rm m0}$ and $n$ and the
right panel shows the observational $\mu(z)$ plot and the
theoretical $\mu(z)$ plot by using the best fit parameters
$\Omega_{\rm m0}=0.46$ and $n=2.6$. }
\end{figure}

In order to see the observational effect of this model, we use the
Wilkinson Microwave Anisotropy Probe (WMAP) data \cite{bennett03}
and the supernova (SN) data compiled by Riess {\it et al.}
\cite{agriess} to fit the model. The parameters $\Omega_{\rm m0}$
and $n$ are determined by minimizing
\begin{equation}
\label{lrmin} \chi^2=\sum_i{[\mu_{\rm obs}(z_i)-\mu(z_i)]^2\over
\sigma^2_i}+{(\mathcal{R}_{\rm obs}-\mathcal{R})^2\over
\sigma^2_\mathcal{R}},
\end{equation}
where $\sigma_i$ is the total uncertainty in the SN observation
and $\sigma_\mathcal{R}$ is the uncertainty in $\mathcal{R}$, the
extinction-corrected distance moduli $\mu(z)=5\log_{10}(d_{\rm
L}(z)/{\rm Mpc})+25$, $d_{\rm L}(z)=c(1+z)\int^{t_0}_{\rm T} d t'/
a(t')$, the CMB shift parameter $\mathcal{R}\equiv
\Omega^{1/2}_{\rm m0}H_0d_{\rm L}(z_{\rm ls})/(c(1+z_{\rm
ls}))=1.716\pm 0.062$ \cite{shift} and $z_{\rm ls}=1089\pm 1$
\cite{bennett03}. The best fit parameters to the gold sample SN
data in Ref.~\refcite{agriess} and the WMAP data are $\Omega_{\rm
m0}=0.46^{+0.09}_{-0.11}$ and $n=2.6^{+4.7}_{-0.5}$ with
$\chi^2=181.4$. The $\Omega_{\rm m0}$ and $n$ 1$\sigma$ confidence
contour is shown in figure \ref{bestfit}. At 99.5\% confidence
level, we have $n\ge 1.7$. By using the best fit parameters
$\Omega_{\rm m0}=0.46$ and $n=2.6$, we find that $\omega_{\rm
Q0}=-2.85$ and the transition redshift $z_{\rm T}=0.44$. From the
dynamical Eq. (\ref{darkom}) of $\omega_{\rm Q}$, we see that
$\omega_{\rm Q}\approx -n\alpha$ when the redshift $z$ is large.
The current matter density and the transition redshift $z_{\rm
T}\sim 0.5$ are consistent with the constraints obtained by some
model independent results \cite{agriess,uavs,daly}.

In conclusion, we propose a new model which explains the past
deceleration and the recent acceleration and recovers the standard
model at early times. In addition, the new model predicts that the
expansion of the universe will slow down in the future.

\section*{Acknowledgments}
The authors thank the anonymous referee for fruitful comments. The
author YG thanks G. Calcagni for pointing out his work. The work
is supported by Chongqing University of Post and Telecommunication
under grants A2003-54 and A2004-05.


\section*{References}
\begin{thebibliography}{sp99}
\bibitem{sp99} S. Perlmutter {\it et al.},
Nature {\bf 391}, 51 (1998); \APJ
 {\bf 517}, 565 (1999); P.M. Garnavich {\it et al.}, \APJ Lett. {\bf 493},
L53 (1998); A.G. Riess {\it et al.}, Astron. J. {\bf 116}, 1009
(1998).
\bibitem{tonry} J.L. Tonry {\it et al.}, \APJ {\bf 594}, 1
(2003); B.J. Barris {\it et al.}, \APJ {\bf 602}, 571 (2004).
\bibitem{raknop03} R.A. Knop {\it et al.},  astro-ph/0309368.
\bibitem{agriess} A.G. Riess {\it et al.},  astro-ph/0402512.
\bibitem{agr} A.G. Riess, \APJ {\bf
560}, 49 (2001).
\bibitem{mstagr}M.S. Turner and A.G. Riess, \APJ {\bf 569}, 18
(2002).
\bibitem{pdb00} P. de Bernardis {\it et al.},
Nature {\bf 404}, 955 (2000); Hanany S {\it et al.}, \APJ Lett.
{\bf 545}, L5 (2000).
\bibitem{bennett03} C.L. Bennett {\it et al.}, \APJ Supp. Ser. {\bf
148}, 1 (2003); D.N. Spergel {\it et al.}, \APJ Supp. Ser. {\bf
148}, 175 (2003).
\bibitem{dark} P.J.E. Peebles and B. Ratra \APJ Lett. {\bf 325}, L17 (1988); B. Ratra and P.J.E. Peebles, \PR D {\bf 37}, 3406 (1988);
C. Wetterich, \NP B {\bf 302}, 668 (1988); P.J.E. Peebles and B.
Ratra, Rev. Mod. Phys. {\bf 75}, 559 (2003).
\bibitem{quint} R.R. Caldwell, R. Dave and P.J. Steinhardt,
\PRL {\bf 80}, 1582 (1998); I. Zlatev, L. Wang and P.J.
Steinhardt, \PRL {\bf 82}, 896 (1999).
\bibitem{pgfmj} P.G. Ferreira and M.
Joyce, \PRL {\bf 79}, 4740 (1997); \PR D {\bf 58}, 023503 (1998);
S. Perlmutter, M.S. Turner and M. White, \PRL {\bf 83}, 670
(1999).
\bibitem{johri} V. Sahni and A.A. Starobinsky, Int. J. Mod. Phys. D {\bf 9}, 373
(2000); C. Rubano and J.D. Barrow, \PR D {\bf 64}, 127301 (2001).
\bibitem{aasss} A.A. Sen and S. Sethi, \PL B {\bf 532}, 159
(2002); Y. Gong, \CQG {\bf 19} 4537 (2002).
\bibitem{uavs} U. Alam, V. Sahni, T.D. Saini and A.A. Starobinsky,
astro-ph/0311364; U. Alam, V. Sahni and A.A. Starobinsky,
astro-ph/0403687.
\bibitem{tachyon}T. Padmanabhan and T.R. Choudhury, \PR D {\bf 66}, 081301
(2002) ; J.S. Bagla, H.K. Jassal and T. Padmanabhan, \PR D {\bf
67}, 063504 (2003); T. Padmanabhan, Phys. Rep. {\bf 380}, 235
(2003); T. Padmanabhan and T.R. Choudhury,  Mon. Not. Roy. Astron.
Soc. {\bf 344}, 823 (2003).
\bibitem{mond} M. Milgrom, Astrophys. J. {\bf 270}, 365 (1983).
\bibitem{mond1}A. Lue and G.D. Starkman, astro-ph/0310005.
\bibitem{brane} P. Bin\'etruy, C. Deffayet and D. Langlois, \NP
B {\bf 565}, 269 (2000); Y. Gong, gr-qc/0005075.
\bibitem{chung} D.J. Chung and K. Freese, \PR D {\bf 61}, 023511
(1999).
\bibitem{freese02} K. Freese and M. Lewis, \PL B {\bf 540}, 1
(2002); K. Freese,  Nucl. Phys. Suppl. Ser. {\bf 124}, 50 (2003);
P. Gondolo and K. Freese, \PR D {\bf 68}, 063509 (2003).
\bibitem{dvali} G.R. Dvali, G. Gabadadze and M. Porrati, \PL B
{\bf 485}, 208 (2000); C. Deffayet, \PL B {\bf 502}, 199 (2001);
C. Deffayet, G.R. Dvali and G. Gabadadze, \PR D {\bf 65}, 044023
(2002); G.R. Dvali and M. Turner, astro-ph/0301510; A. Lue, R.
Scoccimarro and G. Starkman , \PR D {\bf 69}, 044005 (2004).
\bibitem{sen03} S. Sen and A.A. Sen, \APJ {\bf 588}, 1 (2003)
; A.A. Sen and S. Sen, \PR D {\bf 68}, 023513 (2003); M.
Szydlowski and W. Czaja, astro-ph/0309191; W. Godlowski, M.
Szydlowski and A. Karweic, astro-ph/0309569; G. Calcagni,
hep-ph/0402126.
\bibitem{zhu03} Z.H. Zhu and M. Fujimoto, \APJ {\bf 581}, 1
(2003); {\bf 585}, 52 (2003); \APJ {\bf 602}, 12 (2004); Z.H. Zhu
, M. Fujimoto and X.T. He, \APJ {\bf 603}, 365 (2004).
\bibitem{wang} Y. Wang, K. Freese, P. Gondolo and M. Lewis, \APJ {\bf 594}, 25
(2003); T. Multamaki, E. Gaztanaga and M. Manera, Mon. Not. Roy.
Astron. Soc. {\bf 344}, 761 (2003); A. Dev, J.S. Alcaniz and D.
Jain, astro-ph/0305068; W.J. Frith, Mon. Not. Roy. Astron. Soc.
{\bf 348}, 916 (2004); S. Nesseris and L. Perivolaropoulos,
astro-ph/0401556.
\bibitem{gong03} Y. Gong and C.K. Duan,
gr-qc/0311060; astro-ph/0401530; Y. Gong and X.M. Chen,
gr-qc/0402031.
\bibitem{chaply} A. Kamenshchik, U. Moschella and V. Pasquier, \PL B
{\bf 511}, 265 (2001); N. Bilic, G.G. Tupper and R.D. Viollier,
\PL B {\bf 535}, 17 (2002); M.C. Bento, O. Bertolami and A.A. Sen,
\PR D {\bf 66}, 043507 (2002); D. Carturan and F. Finelli, \PR D
{\bf 68}, 103501 (2003); L. Amendola, F. Finelli, C. Burigana and
D. Carturan, J. Cosmology Astroparticle Phys. {\bf 0307}, 005
(2003); J.V. Cunha, J.S. Alcaniz and J.A.S. Lima, \PR D {\bf 69},
083501 (2004).
\bibitem{shift} J.R. Bond, G. Efstathiou and M. Tegmark, Mon. Not. Roy. Astron. Soc. {\bf
291}, L33 (1997); A. Melchiorri, L. Mersini, C.J. \"{O}dman and M.
Trodden, \PR D {\bf 68}, 043509 (2003); Y. Wang and P. Mukherjee,
astro-ph/0312192; Y. Wang and M. Tegmark, astro-ph/0403292.
\bibitem{daly} R.A. Daly and S.G. Djorgovski, \APJ {\bf 597}, 9
(2003); astro-ph/0403664; Y. Gong, astro-ph/0401207.
\end{thebibliography}
\end{document}